\documentclass[prl,twocolumn,amsmath,superscriptaddress,showpacs]{revtex4}
\usepackage{graphicx}
\usepackage{dcolumn}
\usepackage{bm}

\begin{document}

\title{Very large optical rotation generated by Rb vapor in a multi-pass cell.}
\author{S. Li}
\affiliation{Department of Physics,  Princeton University, Princeton, New Jersey 08544, USA}
\affiliation{Department of Measurement Technology and Instruments, Zhejiang University of Science and Technology, Hangzhou 310023, China}
\author{P. Vachaspati}
\affiliation{Department of Physics,  Princeton University, Princeton, New Jersey 08544, USA}
\author{D. Sheng}
\affiliation{Department of Physics,  Princeton University, Princeton, New Jersey 08544, USA}
\author{ N. Dural}
\affiliation{Department of Physics,  Princeton University, Princeton, New Jersey 08544, USA}
\author{M. V. Romalis}
\affiliation{Department of Physics,  Princeton University,
Princeton, New Jersey 08544, USA}

\begin{abstract}
Paramagnetic Faraday rotation is a powerful technique for atom sensing widely used in quantum non-demolition measurements, fundamental symmetry tests, and other precision measurements. We demonstrate the use of a multi-pass optical cell for Faraday rotation spectroscopy and observe polarization rotation in excess of 100 radians from spin-polarized Rb vapor. Unlike optical cavities, multi-pass cells have a deterministic number of light passes and can be used to measure large optical rotations. We also observe a 10-fold suppression of transverse spin relaxation when Rb atoms are placed in a coherent superposition state immune to spin-exchange collisions.

\end{abstract}


\maketitle

Polarization rotation of an off-resonant linearly polarized light is a simple method for measuring dispersive atom-light interactions \cite{Budker0}. It is minimally destructive for atom coherences and therefore is widely used for atom probing in the most demanding applications, including quantum-non-demolition measurements \cite{Takahashi,Bigelow,Takahashi1,Shah,Mitchell,Polzik},  tests of fundamental symmetries \cite{Fortson1,Fortson2,Brown}, and optical magnetometry \cite{Kominis,Budker}. The size of the Faraday rotation signal is proportional to the optical path length and can be increased in multi-pass geometries. Optical cavities have been previously used to amplify optical rotation in fundamental physics experiments \cite{Zavattini} as well as in quantum non-demolition measurements \cite{Vuletic,Thompson}.

Optical cavities, however, are difficult to use for measurements of large dispersive interactions. The cavity finesse drops when the total rotation angle approaches one radian and the cavity resonance splits into separate resonances for $\sigma^+$ and $\sigma^-$ light. As a result, laser frequency scanning techniques are typically used for measurements of strong atom-cavity interactions \cite{Kimble,Essinger,Thompson}. Here we demonstrate the use of a multi-pass cell \cite{White,Herriott,Silver} for Faraday rotation spectroscopy. Unlike optical cavities, multi-pass cells have a deterministic path length for every photon and no optical resonances, which allows direct measurement of optical rotation of an arbitrary size. We observe Faraday rotation in excess of 100 radians from spin-polarized Rb atoms. The periodic signal of a balanced polarimeter for large rotation angles allows particularly accurate measurement of the rotation amplitude by counting the number of complete $\pi$ rotations.

Multi-pass cells have a number of other advantages for atom sensing compared to optical cavities. The mode volume of an optical cavity is naturally small unless one uses flat or nearly flat mirrors with stringent surface quality requirements similar to that of an etalon because it is difficult to maintain light wavefront flatness of much better than the wavelength of light on a centimeter scale. As a result, only a relatively small number of atoms can typically interact with an optical cavity, reducing the fundamental sensitivity of precision measurements. Unlike an optical cavity, there are no standing waves in the multi-pass cell, the light beam never retraces itself and interacts mostly with different atoms on each pass. Hence a multi-pass cell has a larger effective interaction volume, can operate at any laser frequency within mirror reflectivity range, has 100\% power coupling efficiency, and accepts beams with a range of sizes. The alignment tolerances are also significantly relaxed compared to a cavity, since the relevant length scale is the beam size instead of the wavelength of light.  The robustness of multi-pass cells has allowed us to place mirrors inside a sealed glass cell filled with alkali-metal and obtain more than 100 passes through the cell without any active adjustment.  Multi-pass cells have been operated with up to five hundred passes \cite{Wilson}. The number of passes is fundamentally limited by the ratio of the mirror area to the beam area.

 Using the multi-pass cell we also observe a large reduction of transverse spin relaxation rate when Rb atoms are prepared in a coherent superposition state  that is immune to spin-exchange relaxation due to nearly complete spin polarization in the rotating frame. Such state represents a simple example of decoherence-free subspace \cite{Chuang} and can be used to improve long-term frequency resolution with quantum non-demolition measurements \cite{Vasilakis}.

Multi-pass cells have a long history in optical absorption spectroscopy of weak atomic transitions or trace gases \cite{White,Herriott}, but, to our knowledge, have not been previously used for optical rotation spectroscopy.  Here we use a dense-beam design with two cylindrical mirrors and a hole at the center of one of the mirrors for entrance and exit beams \cite{Silver}, as shown in Fig.~1.  The axes of curvature of the two mirrors are rotated relative to each other so that laser reflections form a 2-dimensional Lissajous pattern on mirror surfaces. The number of passes is determined solely by the mirror rotation angle $\theta$ and the distance between mirrors $d$.  The mirrors had a curvature of 10 cm,  diameter of 12 mm, hole diameter of 1.5 mm, and distance between mirrors $d=2.3$ cm. The mirrors are initially mounted on 6-axes adjustment stages and a multi-pass pattern with a clear exit beam is established. Different types of beam patterns can be obtained depending on the distance between mirrors and their relative rotation angle, we selected a pattern with a relatively uniform density of light beams on the mirrors, as shown in the inset of Fig.~1. The number of passes is equal to $102\pm 4$ as determined by counting the number of spots on the mirrors. The mirrors are then cemented to a glass tube which is installed inside a glass cell. We use Al$_2$O$_3$-based cement and an outer dielectric layer of Al$_2$O$_3$ on the mirror coatings to increase their resistance to the alkali-metal vapor.

\begin{figure}
\centering
\includegraphics[width=8.5cm]{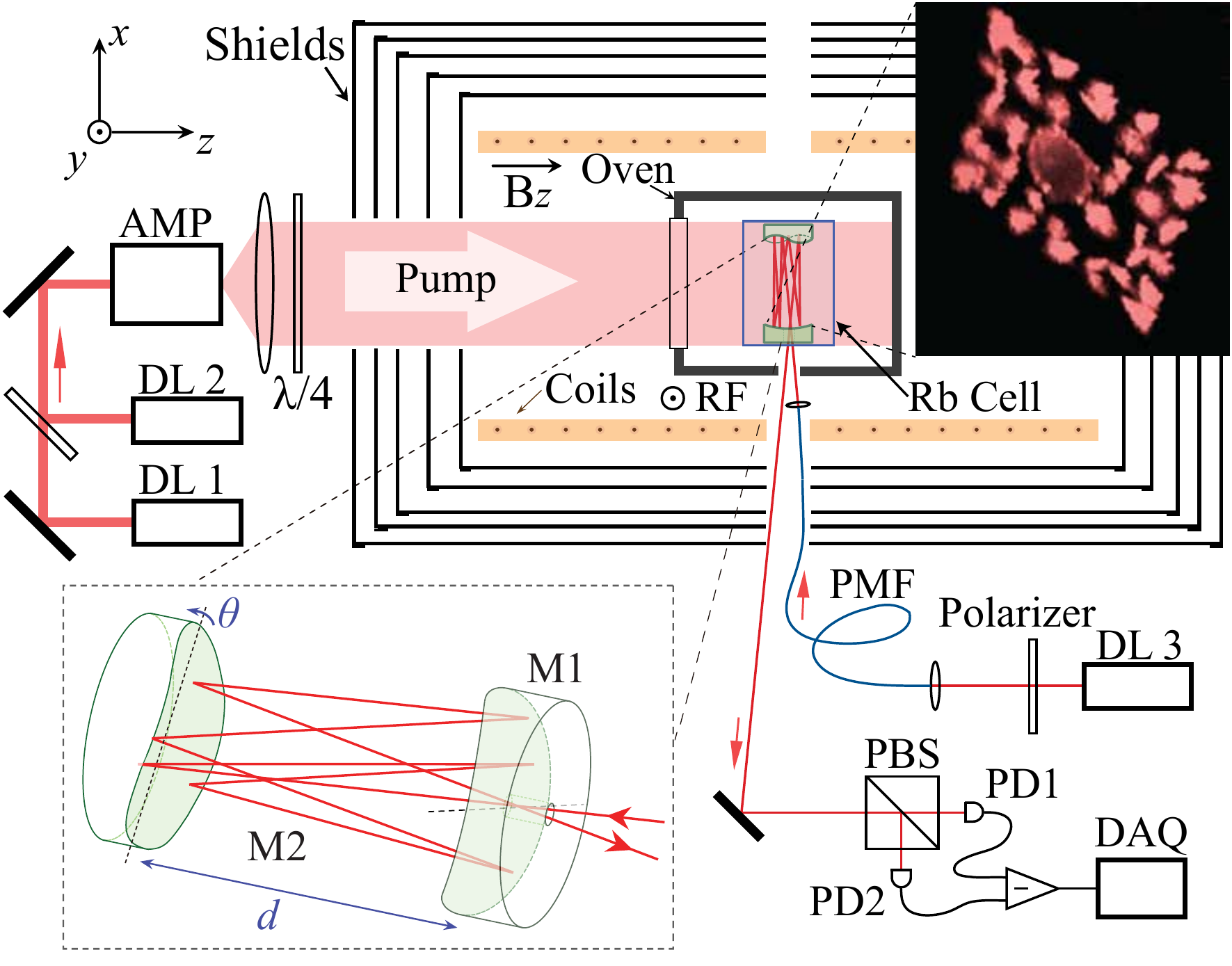}
\caption{Schematic of the experimental apparatus and the multi-pass cell, the inset shows a photo of the beam spots on the entrance mirror. }
\end{figure}

The cell is filled with 70 torr of N$_2$ and a drop of $^{87}$Rb enriched to 99\%. It is heated to 133$^{\circ}$C a non-magnetic oven by  high frequency AC currents such that the mirrors inside the cell are maintained at a higher temperature than the cell walls to prevent Rb condensation. Using Rb-Rb spin-exchange cross-section measured in \cite{Walter}, we determine the density of Rb from the rate of transverse spin relaxation at low Rb polarization to be $(3.0\pm0.15)\times10^{13}$ cm$^{-3}$.  The cell is placed in a five-layer magnetic shield with inner coils that can generate uniform magnetic fields along each of the three orthogonal directions. In order to achieve a high degree of polarization in Rb vapor we use two distributed Bragg reflection (DBR) lasers
tuned to the two ground state hyperfine lines of the D1 transition in $^{87}$Rb. The laser beams are combined into  one beam on a non-polarizing
beamsplitter and are amplified by a tapered laser amplifier to a power of 500 mW. The laser beam is then circularly polarized and expanded to fully illuminate the region between mirrors. The probe beam is generated by another DBR laser and coupled to a single mode PM fiber with an output collimator  mounted near the multi-pass cell for stable beam injection into the cell.  The probe power entering the cell is about 0.5 mW and is collimated to a beam diameter of about 0.7 mm to pass through the entrance hole in the first mirror.  The probe frequency is detuned to the blue side of the D1 resonance since the absorption cross-section is slightly smaller on the blue wing of the Lorentzian \cite{Ottinger,Romalis}. For detuning of  $106$ GHz to the blue from the $F=2$ line the probe transmission is equal to approximately 25\% of the transmission for very far detuning with little atomic absorption.  After exiting from the cell, the probe beam polarization is analyzed by a balanced polarimeter oriented at 45$^{\circ}$ to the initial probe  polarization. The optical signals are detected by low-capacitance Si photodiodes and digitized at 200 MHz by a 16-bit acquisition system.
\begin{figure}
\centering
\includegraphics[width=8.6cm]{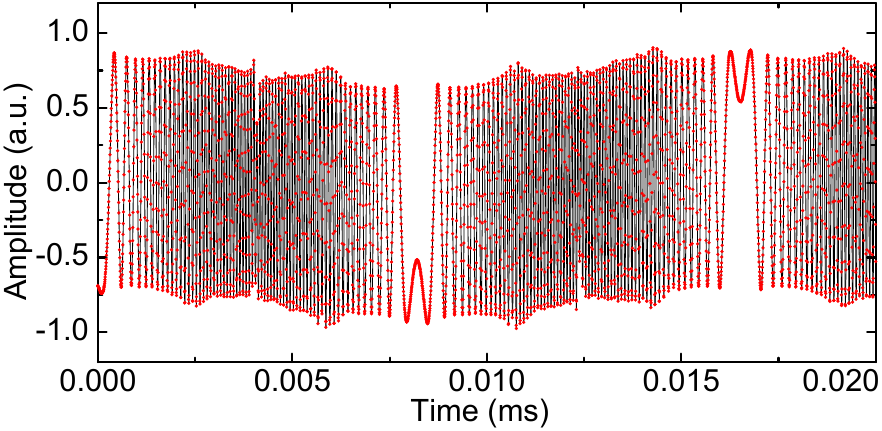}
\caption{Signal recorded by the balanced polarimeter immediately after the $\pi/2$ pulse.}
\end{figure}

The atoms are optically pumped along a constant magnetic field $B_{z}$ applied parallel to the pump beam. The pump beam is turned off and a $\pi/2$ pulse for $^{87}$Rb atoms at 60 kHz
is applied along the $\hat{y}$ direction. The amplitude of the $\pi/2$ pulse is adjusted so that it takes 3 full periods, which results in approximately $2\pi$ rotation for $^{85}$Rb atoms, further reducing their small contribution to the signal. Fig.~2 shows the polarimeter signal recorded after the $\pi/2$ pulse. The complicated structure of the signal is due to optical rotation far exceeding $\pm \pi/4$ monotonic range of the balanced polarimeter. The signal $ V_{\rm out}$ generated by Faraday rotation due to $^{87}$Rb Larmor precession after the pulse is given by
\begin{eqnarray}
V_{\rm out}&=&V_1-V_2=V_0\sin(2\phi)  \\
\phi&=&\phi_0 \sin(\omega_{L}t)\exp(-t/T_{2})
\end{eqnarray}
where $V_0$ is the maximum voltage corresponding to full probe intensity. The maximum Faraday rotation angle $\phi_0$ in the regime of far detuning is given by
\begin{equation}
\phi_{0}= \frac{1}{2} n l r_{e} c P f /(\nu-\nu_{0})  \label{rot}
\end{equation}
where $r_{e}=2.82\times10^{-13}$ cm is the classical electron radius, $P$ is the initial transverse spin polarization of Rb atoms,
$f=0.342$ is the oscillation strength and $\nu-\nu_{0}$ is the detuning from the D1 $F=2$ resonance.
The large optical path length $l$, equal to 2.4 meters for data in Fig.~2, results in multiple "wraping" of the polarimeter output. The periodicity of the polarimeter signal can be used to increase the absolute accuracy of optical rotation measurements and alleviate practical limitations due to noise and finite dynamic range of experimental measurements by counting the number of $\pi$ rotations, similar to flux-counting techniques used in Superconducting Quantum Interference Devices \cite{Faley}.

We developed an algorithm to restore the absolute rotation
angle from the polarimeter output data. The algorithm corrects for small distortions of the data by calibrating the signal between each maximum and minimum to correspond to a $\pi/2$ rotation, with special treatment for the turning points of the rotation angle. The results of this "unwrapping" are shown in Fig. 3(a). The optical rotation amplitude reaches 104 radians immediately after the RF pulse. Eq.~(\ref{rot}) predicts maximum rotation of $100\pm6$ radians for full spin polarization after applying a small correction for the rotation due to the D2 line.

\begin{figure}
\centering
\includegraphics[width=8.5cm]{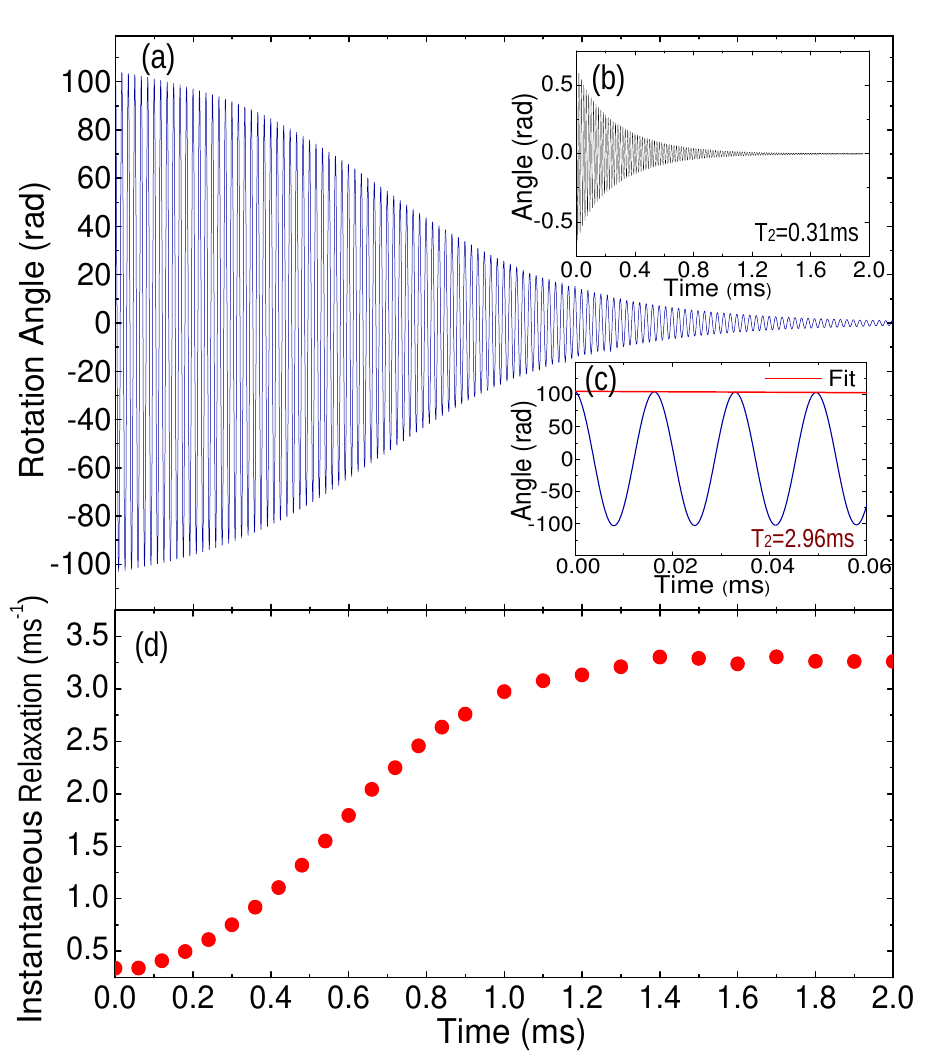}
\caption{"Unwrapped" optical rotation signal for large initial polarization (a) and  small initial polarization (inset b). The instantaneous transverse spin relaxation rate is shown in (d), determined from decay of a few periods as shown in inset (c). }
\end{figure}
The decay of the transverse spin polarization is non-exponential, as seen in Fig.~3a due to the effects of spin-exchange relaxation. It has been shown previously \cite{Happer,RF} that spin-exchange relaxation is suppressed when all alkali-metal atoms are pumped into a stretched state with $F=2$, $m=2$. In our case, after the RF pulse the atoms are in a coherent superposition of all $m$ states in the $F=2$ hyperfine manifold. However, in the rotating frame the atoms remain in the stretched state with maximum spin polarization. Since spin-exchange interactions are rotationally invariant,  such coherent state remains immune from spin-exchange relaxation.  In contrast, if the initial spin polarization is small then the transverse relaxation is exponential, as shown in Fig. 3(b). We calculate the instantaneous rate of spin relaxation as illustrated in Fig. 3(c) and plot this time-dependent decoherence rate in Fig. 3(d). As expected, the relaxation rate is smallest immediately after the RF pulse, a factor of 10 smaller than the relaxation rate at small polarization, which implies about 94\% initial spin polarization \cite{Smullin}. The initial  polarization is limited by quick relaxation of atoms near mirror surfaces. Such non-exponential relaxation is a crucial ingredient for improving long-term resolution of frequency measurements using spin squeezing techniques \cite{Vasilakis}.

\begin{figure}
\centering
\includegraphics[width=8.5cm]{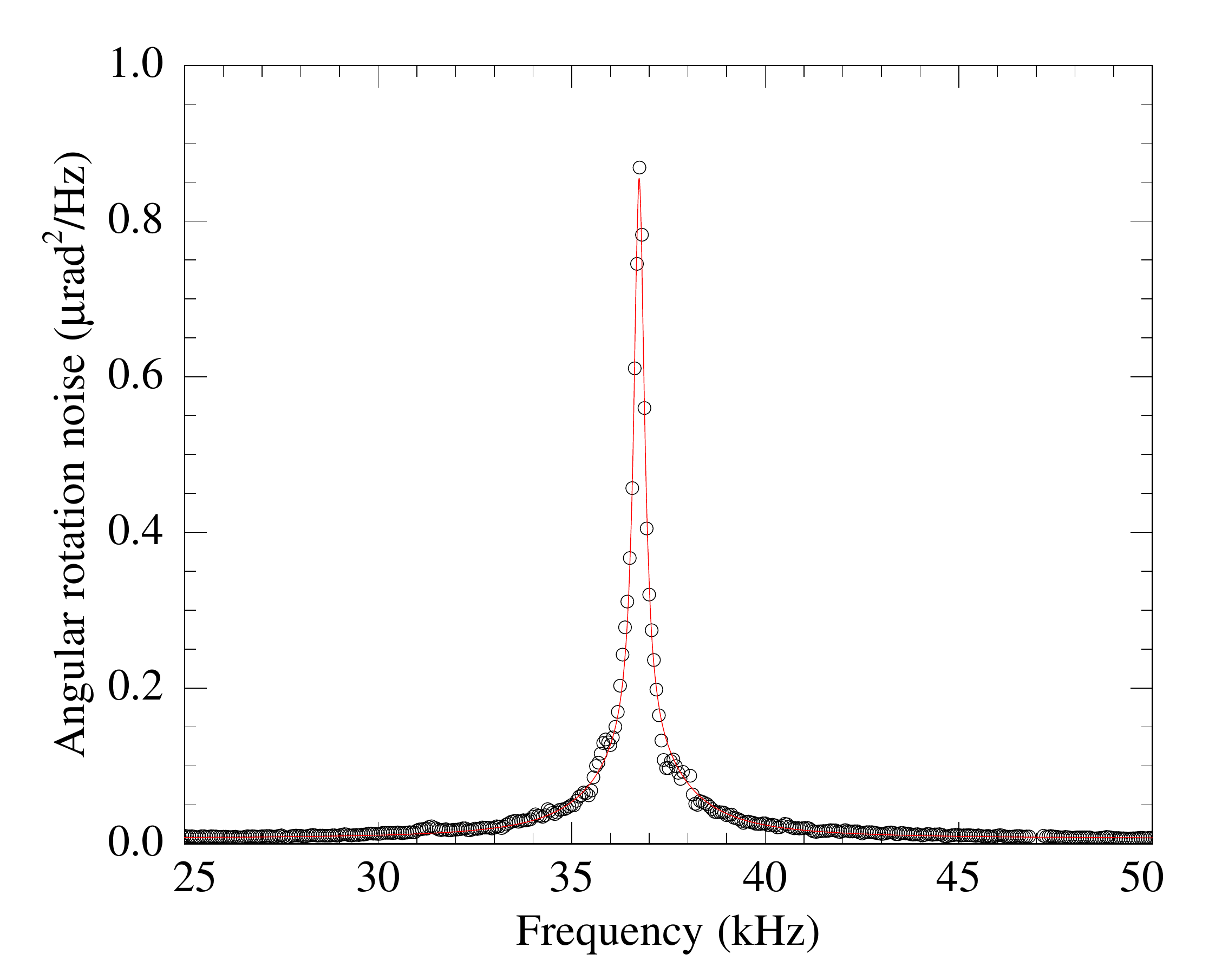}
\caption{Spectrum of atomic spin noise for unpolarized Rb atoms. For this measurement we used a multi-pass cell with 112 passes, the density of Rb was $1\times 10^{13}$cm$^{-3}$ and the Rb Larmor frequency was equal to 37 kHz. }
\end{figure}

Another advantage of the multi-pass geometry is the increase in the contrast of atom shot noise relative to the photon shot noise. In the regime where the probe scattering rate is dominating spin relaxation, the ratio of atom shot noise to photon shot noise is proportional to the optical density on resonance \cite{Shah}. Hence in a multi-pass cell the atom shot noise contrast  can be significantly increased.  A plot of atomic noise spectrum for unpolarized Rb atoms is shown in Fig.~4. The contrast ratio of atomic to photon shot noise is equal to 130, much larger than in  previous experiments utilizing a single-pass arrangement \cite{Crooker,Kominis1,Shah}.  The shape of the noise peak is quite different from a single Lorentzian due to the effects of diffusion \cite{Walsworth}. In multi-pass cells the beam size has significant variation within the cell. In addition, some atoms can diffuse from one beam path to another over their coherence time. As a result, the effective measurement time for atoms has a wide distribution, which results in a distribution of effective linewidths. Phenomenologically we  quantify the diffusion effect by fitting the spectrum in Fig.~4  to a sum of two Lorentzians. The narrow Lorentzian component has a half-width of 170 Hz, consistent with the spin-exchange linewidth, indicating that for some atoms diffusion does not limit the measurement time.  The broad component has a linewidth of 1.3 kHz, indicating the short timescale of atomic diffusion for a single beam. The areas of the two Lorentzian components are approximately equal.  Higher buffer gas pressure can be used to confine the atoms into a single probe beam for the duration of their coherence time. Alternatively, in  the absence of the buffer gas the atoms can sample many laser beams as long as the walls of the cell have a good anti-relaxation coating \cite{Seltzer,Ledbetter}. It remains a challenge to make an anti-relaxation coating with very low light absorption and scattering that would be suitable for multi-pass cells.

In summary, we introduced  optical rotation spectroscopy technique in a multi-pass cell and pointed out its advantages over optical cavities. We observed optical rotation signals in excess of 100 radians from polarized Rb vapor. We also measured quantum spin noise with a large contrast ratio and observed a large suppression of spin-exchange relaxation in a coherent superposition of $F=2$ states in $^{87}$Rb vapor, demonstrating all ingredients necessary to achieve an improvement in {\it long-term} frequency resolution using spin squeezing techniques \cite{Vasilakis}. This work also illustrates general advantages of multi-pass cell geometry for interrogation of a large number of atoms at high optical density. In addition to the Faraday rotation measurements, it will be applicable to many other atomic experiments benefiting from a large optical depth, such as  electromagnetically induced transparency (EIT)\cite{Harris}, slow and stopped light \cite{Hau}, and four-wave mixing \cite{Harris1}. We acknowledge financial support from DARPA.

\end{document}